\documentclass[journal]{IEEEtran}

\usepackage{booktabs}
\usepackage{stfloats}

\ifCLASSINFOpdf
   \usepackage[pdftex]{graphicx}
   \DeclareGraphicsExtensions{.pdf,.jpeg,.png}
\else
\fi

\usepackage[ruled,linesnumbered]{algorithm2e}

\usepackage{amsmath,amsfonts}
\usepackage{amssymb}
\usepackage{nicefrac}

\usepackage{color}
\usepackage{subcaption}

\usepackage[makeroom]{cancel}

\usepackage{multirow}

\usepackage{mathrsfs}

\usepackage{soul}

\usepackage{bm}
\usepackage{caption}

\DeclareCaptionLabelSeparator{colquad}{.\quad}

\begin{document}

\title{Online Detection of Low-Quality Synchrophasor Data Considering Frequency Similarity}

\author{Wenyun~Ju,~\IEEEmembership{Member,~IEEE}, Horacio Silva-Saravia,~\IEEEmembership{Member,~IEEE}, Neeraj Nayak,~\IEEEmembership{Member,~IEEE}, Wenxuan Yao,~\IEEEmembership{Member,~IEEE}, Yichen Zhang,~\IEEEmembership{Member,~IEEE}, Qingxin  Shi,~\IEEEmembership{Member,~IEEE}, Fan Ye,~\IEEEmembership{Member,~IEEE} 
\\

%\thanks{

%W. Ju, H. Silva-Saravia and N. Nayak are with Electric Power Group, LLC, Pasadena, CA 91101 USA. W. Yao is with University of Tennessee, Knoxville, TN 37996 USA. Y. Zhang is with Argonne National Laboratory, Lemont, IL 60439 USA. F. Ye is with GE Digital, Bothell, WA 98011 USA.
\vspace{-0.2cm}
%}
\vspace{-0.35cm}
}
\maketitle
\begin{abstract}
This letter proposes a new approach for online detection of low-quality synchrophasor data under both normal and event conditions. The proposed approach utilizes the features of synchrophasor data in time and frequency domains to distinguish multiple regional PMU signals and detect low-quality synchrophasor data. The proposed approach does not require any offline study and it is more effective to detect low-quality data with apparently indistinguishable profiles. Case studies from recorded synchrophasor measurements verify the effectiveness of the proposed approach.
\end{abstract}

\begin{IEEEkeywords}
Synchrophasor measurements, low-quality synchrophasor data, frequency domain, data analytics. 
\end{IEEEkeywords}

\IEEEpeerreviewmaketitle

\section{Introduction}

Low-quality synchrophasor data are widely seen in practice. It represents data that cannot accurately reflect the underlying system behavior \cite{FA1}. Due to the inherent networked electrical couplings between individual buses, synchrophasor data from regional PMU signals generally have similar dynamic behaviors in both normal and event operating conditions \cite {FA1,FA2}. This is called strong spatial-temporal correlation. This correlation becomes relatively weak if data anomalies exist. The detection of spatial-temporal anomalies including random spikes, repeated data and false data injection is the objective of this letter. Other types of data anomalies such as missing data and high sensing noises are not considered by this letter. 

%Communication problems can induce data dropout or temporary data unavailability, they can be flagged by NaN or zero values. The anomalies caused by high sensing noises can be identified by signal processing tools. These types of data anomalies are not considered in this paper.

In recent years, model-free based methods \cite{FA1,FA2,FA3,FA4,FA5} are exploited to achieve more reliable detection under inaccurate topology information or parameter errors. Reference \cite{FA3} proposes a method to identify and correct low-quality data based on low-rank property of the Hankel structure. However, its complicated optimizations make it hard for online application. References \cite{FA4,FA5} use machine learning methods to detect low-quality data and require time-consuming labelled dataset for training. Reference \cite{FA1} proposes a density-based local outlier approach to detect low-quality data. It requires high-quality historical database for multiple PMU signals. Reference \cite{FA2} proposes an approach based on spatial-temporal nearest neighbor (STNN) discovery. Some of them are not capable to detect low-quality data that appears in the same time period of multiple regional PMU signals \cite{FA2,FA4,FA5}.

%These methods require either large computational burden such as nonlinear %optimizations [3] or time-consuming offline study such as labelled dataset for %training [4],[5] and high-quality historical database [1].

This letter addresses issues memtioned above by developing a model-free approach which utilizes the features of synchrophasor data in time and frequency domains for detecting low-quality synchrophasor data online. The advantages are:

\begin{itemize}
      
      \item It does not require any offline study and training and few online computational efforts are required.
      
      \item It is more effective to detect low-quality synchrophasor data with
      apparently indistinguishable profiles.
      
      \item It can differentiate event data from low-quality data.
      
      \item The threshold to detect low-quality synchrophasor data is meaningful and it is much easier to understand and set.  
           
\end{itemize} 
\vspace{-0.18cm}
\section{Quantifying Similarity for Two Data Curves}

The synchrophasor data matrix $M$ from $N$ regional PMU signals is divided into $T$ time periods. Let $M_{i}$ and $M_{j}$ denote the data for one same time period of signals $i$ and $j$. Three indices are used to comprehensively quantify the similarity between $M_{i}$ and $M_{j}$ from both time and frequency domains.
\vspace{-0.3cm}
\subsection{Dynamic Change Similarity}
Define Dynamic Change Similarity as (1), denoted by ${I}_{dcs}$. 
\begin{eqnarray}
&& {I}_{dcs} = e^{1-\gamma} , \gamma = max[\frac{\sigma_{i}}{\sigma_{j}},\frac{\sigma_{j}}{\sigma_{i}}]
\end{eqnarray}
where $\sigma_{i}$ and $\sigma_{j}$ are the standard deviations of $M_{i}$ and $M_{j}$.

The range of ${I}_{dcs}$ is [0, 1]. The closer ${I}_{dcs}$ is to 1, the more similar of $M_{i}$ and $M_{j}$ in terms of strength of dynamic change.
\vspace{-0.8cm}
\subsection{Frequency Magnitude Similarity}

The frequency response of a system with $M_{i}$ as the input and $M_{j}$ as the output is:
\begin{eqnarray}
&& H(f) = \frac{|M_{j}(f)|}{|M_{i}(f)|}e^{j(\phi_{M_{j}}(f)-\phi_{M_{i}}(f))}
\end{eqnarray}
where $M_{i}(f)$ and $M_{j}(f)$ represent the Fourier transforms (FTs) of $M_{i}$ and $M_{j}$. $|M_{i}(f)|$ and $|M_{j}(f)|$ are the frequency magnitudes of the FTs of $M_{i}$ and $M_{j}$. $\phi_{M_{i}}(f)$ and $\phi_{M_{j}}(f)$ are the phase angles of the FTs of $M_{i}$ and $M_{j}$.

$|H(f)|$ equals to 1 for all the values of $f$ if $M_{i}$ and $M_{j}$ are the same. Eqn. (3) is used for frequency magnitude and it maps all the values to the range of [0, 1] for each frequency.

\begin{eqnarray}
&& S(f) = 1-tanh(\frac{|20log_{10}|H(f)||}{\lambda})
\end{eqnarray}
where $\lambda$ is a sensitivity parameter for magnitude distance.

For a frequency range with $m$ frequency values, Frequency Magnitude Similarity \cite{FA6}, denoted by $I_{fms}$, is:
\begin{eqnarray}
&& I_{fms} = \frac{\sum_{k=1}^{m}S(f_{k})}{m}
\end{eqnarray}

\subsection{Frequency Phase Similarity}
The angle of frequency response $H(f)$, denoted by $\phi(f)$, equals to zero for all values of frequency if $M_{i}$ and $M_{j}$ are the same. Phase angle distance is defined for each frequency:
\begin{eqnarray}
&& A(f) = 1-tanh(\frac{1}{2\pi}\frac{|\phi(f)|}{\epsilon})
\end{eqnarray}
where $\epsilon$ is a sensitivity parameter for angle distance.

For a given frequency range with $m$ frequency values, Frequency Phase Similarity \cite{FA6}, denoted by $I_{fps}$, is:
\begin{eqnarray}
&& I_{fps} = \frac{\sum_{k=1}^{m}A(f_{k})}{m}
\end{eqnarray}

Note that $I_{dcs}$ can quantify the similarity of $M_{i}$ and $M_{j}$ in the time domain. $I_{fms}$ and $I_{fps}$ quantify the similarity of $M_{i}$ and $M_{j}$ from frequency domain. $M_{i}$ and $M_{j}$ have different dynamic behaviors if either one has low-quality data, $I_{dcs}$, $I_{fms}$ and $I_{fps}$ values tend to be close to 0. Otherwise, $I_{dcs}$, $I_{fms}$ and $I_{fps}$ values tend to be close to 1.

We use a simple example in Fig. 1 to illustrate the necessity for considering $I_{fms}$ and $I_{fps}$ to distinguish two data curves. 
\begin{figure}[!htbp]
\centering
\vspace{-0.7cm}
\includegraphics[width=43mm]{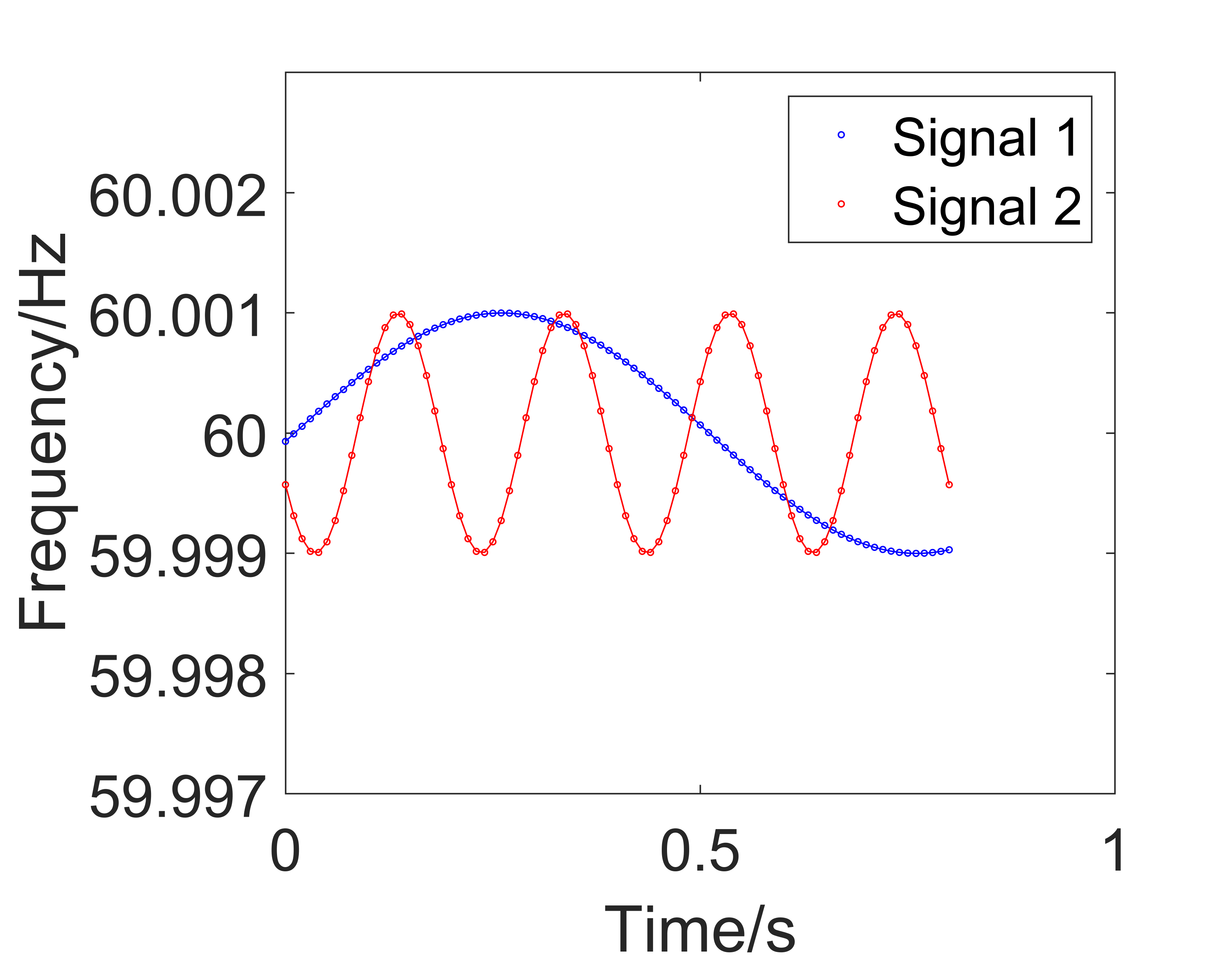}
\caption{An example for two signals.}
\label{Fig1}
\vspace{-0.4cm}
\end{figure}

The standard deviations of signals 1 and 2 are almost the same, the difference is only 0.0001\%. Therefore the LOF approach will conclude that they have the same dynamic change and there is no difference between them. However, the values of $I_{fms}$ and $I_{fps}$ are calculated as 0.4355 and 0.6615, they quantify the difference between signals 1 and 2 more accurately in the frequency domain. 

\section{Proposed Approach for Detecting Low-Quality Synchrophasor Data}
\vspace{-0.10cm}
\subsection{Similarity Degree}
By weighting $I_{dcs}$, $I_{fms}$ and $I_{fps}$, we can have ${I}_{sd}^{ij}$
\begin{eqnarray}
&& {I}_{sd}^{ij} = \omega_{1}I_{dcs}+ \omega_{2} I_{fms} + \omega_{3} I_{fps}
\end{eqnarray}
where $\omega_{1}$, $\omega_{2}$ and $\omega_{3}$ are the weights. $\omega_{1}$ + $\omega_{2}$ + $\omega_{3}$ = 1.

For the $i$th PMU signal at the $k$th time period, calculate the ${I}_{sd}^{ij}$ for each pair of data curves $M_{i}$ and $M_{j}$ to obtain the set of $\{I_{sd}^{i1},...,I_{sd}^{ij},...,I_{sd}^{iN}\}$, then calculate the mean value to obtain the Similarity Degree, denoted by $I_{sd}^{i}$.
\begin{eqnarray}
&& I_{sd}^{i} = \frac{\sum_{j=1,j\neq i}^{N}I_{sd}^{ij}}{N-1}   \quad i=1,...,N
\end{eqnarray}
where $N$ is the number of regional PMU signals. The frequency characteristics of low-quality synchrophasor data especially random spikes and false data injection in multiple regional PMU signals are expected to be different, $I_{sd}^{i}$ is more accurate to distinguish low-quality synchrophasor data in multiple regional PMU signals.
\vspace{-0.10cm}
\subsection{Proposed Approach for Detecting Low-Quality Data}
With the set of $\{I_{sd}^{1},...,I_{sd}^{i},...,I_{sd}^{N}\}$, for the $i$th PMU signal, it is detected as a candidate PMU signal with low-quality data at the $k$th time period if its value of $I_{sd}^{i}$ satisfies (9). 
\begin{eqnarray}
&& I_{sd}^{i}<\zeta   \quad i=1,...,N
\end{eqnarray}
where $\zeta$ is the threshold. The range of $I_{sd}^{i}$ is [0, 1]. The closer $I_{sd}^{i}$ is to 0, the more dissimilar of the $i$th PMU signal compared with other PMU signals. Therefore $\zeta$ is meaningful and it is easier to understand and tune for online detection.

The algorithm for detection of low-quality synchrophasor data can be summarized as:

Step 1: Obtain the synchrophasor data matrix $M$ of $N$ regional PMU signals for the $k$th time period. 

Step 2: For the $i$th ($i$=1,...,$N$) PMU signal, calculate the ${I}_{sd}^{ij}$ for each pair of data curves $M_{i}$ and $M_{j}$ using (1)-(7) and obtain $\{I_{sd}^{i1},...,I_{sd}^{ij},...,I_{sd}^{iN}\}$.

Step 3: Calculate $I_{sd}^{i}$ for the $i$th ($i$=1,...,$N$) PMU signal using (8).

Step 4: Identify the candidate PMU signals with low-quality synchrophasor data using (9), flag the data points in the $k$th time period of the candidate PMU signals as low-quality data, then go to the ($k+1$)th time period.

\section{Case Studies}

Case studies use recorded PMU measurements from normal and event conditions from utilities. $\lambda$ is given as 10, and $\epsilon$ is set as 0.5. They are the same with \cite{FA6}. The frequency range is [0, 5] Hz and higher frequency band is not considered. $\omega_{1}$, $\omega_{2}$, $\omega_{3}$ and $\zeta$ are set as 0.3, 0.35, 0.35 and 0.3, respectively. The LOF approach \cite{FA1} is implemented for comparison. The similarity metric with low variance $f_{L}$ is used and the LOF threshold is set as 10 (100 in \cite{FA1}). We give a smaller value for the LOF threshold in order to achieve better performance for the LOF approach. The moving window length is 80 data points for two approaches. Synchrophasor measurements within the current moving window are identified to contain low-quality data if there are already 15 consecutive moving windows prior to this current window detected in order to avoid the false alarms \cite{FA1}.

\subsection{Scenario 1: Detection of indistinguishable low-quality synchrophasor data under normal condition}
The synchrophasor data of 22 frequency signals ($f$1,$f$2,...,$f$22) recorded from normal condition are used. Among them, $f$10 contains random spikes and repeated data.
\begin{figure}[!htbp]
\vspace{-0.5cm}
\centering
\subfloat[Proposed approach]{\includegraphics[width=46.5mm]{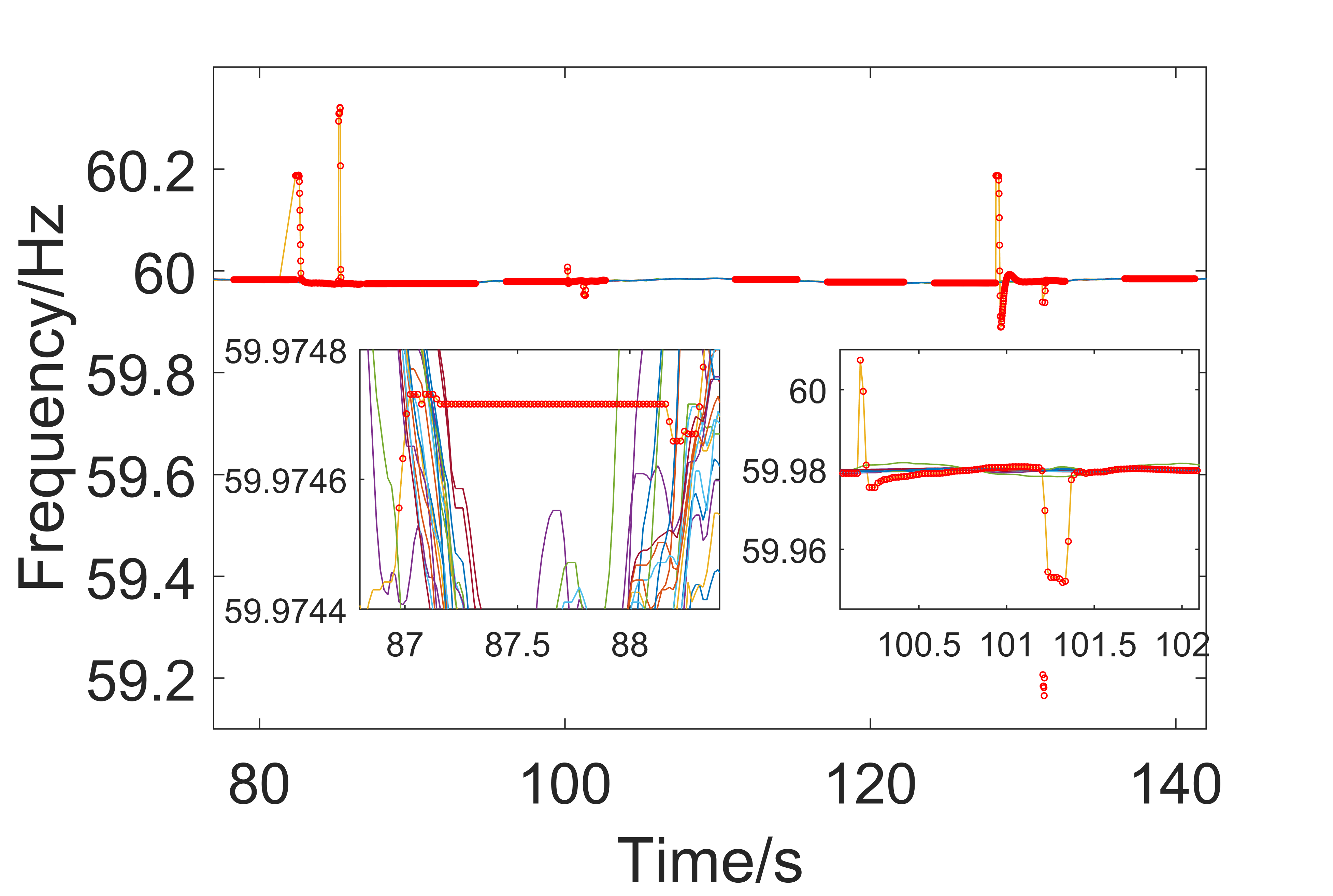}}
\subfloat[LOF approach]{\includegraphics[width=46.5mm]{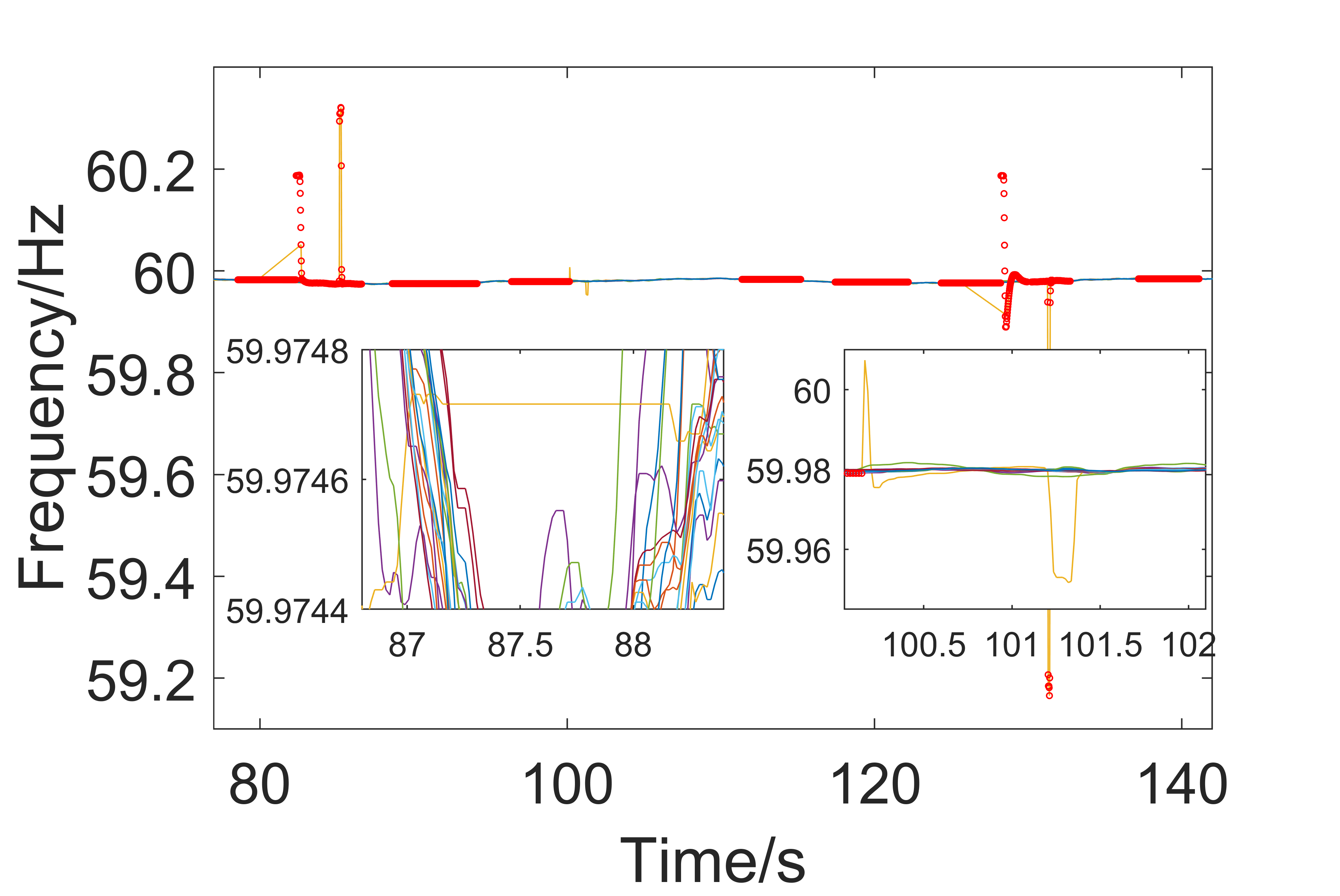}} 
\caption{Identified random spikes and repeated data for Scenario 1.}
\vspace{-0.3cm}
\label{fig.EcUND}
\end{figure}

The data points detected as low-quality synchrophasor data using the proposed and LOF approaches are marked by red in Fig. 2(a) and Fig. 2(b). They are almost the same except that the proposed approach detects additional data points highlighted by red in the zoomed subfigures in Fig. 2(a) and the LOF approach fails to detect them. The reason is further investigated as below.
\begin{figure}[!htbp]
\centering
\vspace{-0.2cm}
\subfloat[$f$1 vs other signals ($f$10 excluded)]{\includegraphics[width=46.5mm]{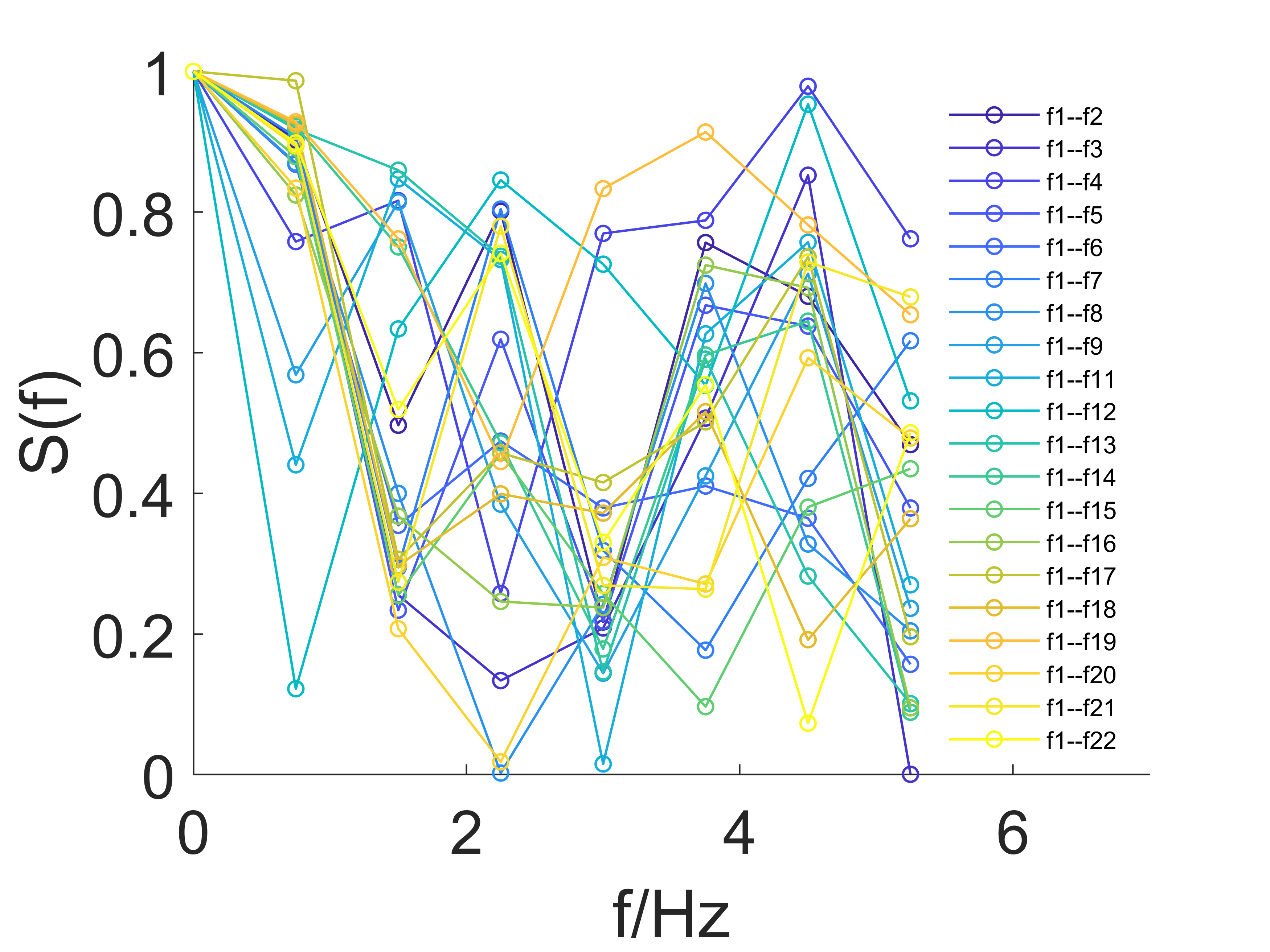}}
\subfloat[$f$10 vs other signals]{\includegraphics[width=46.5mm]{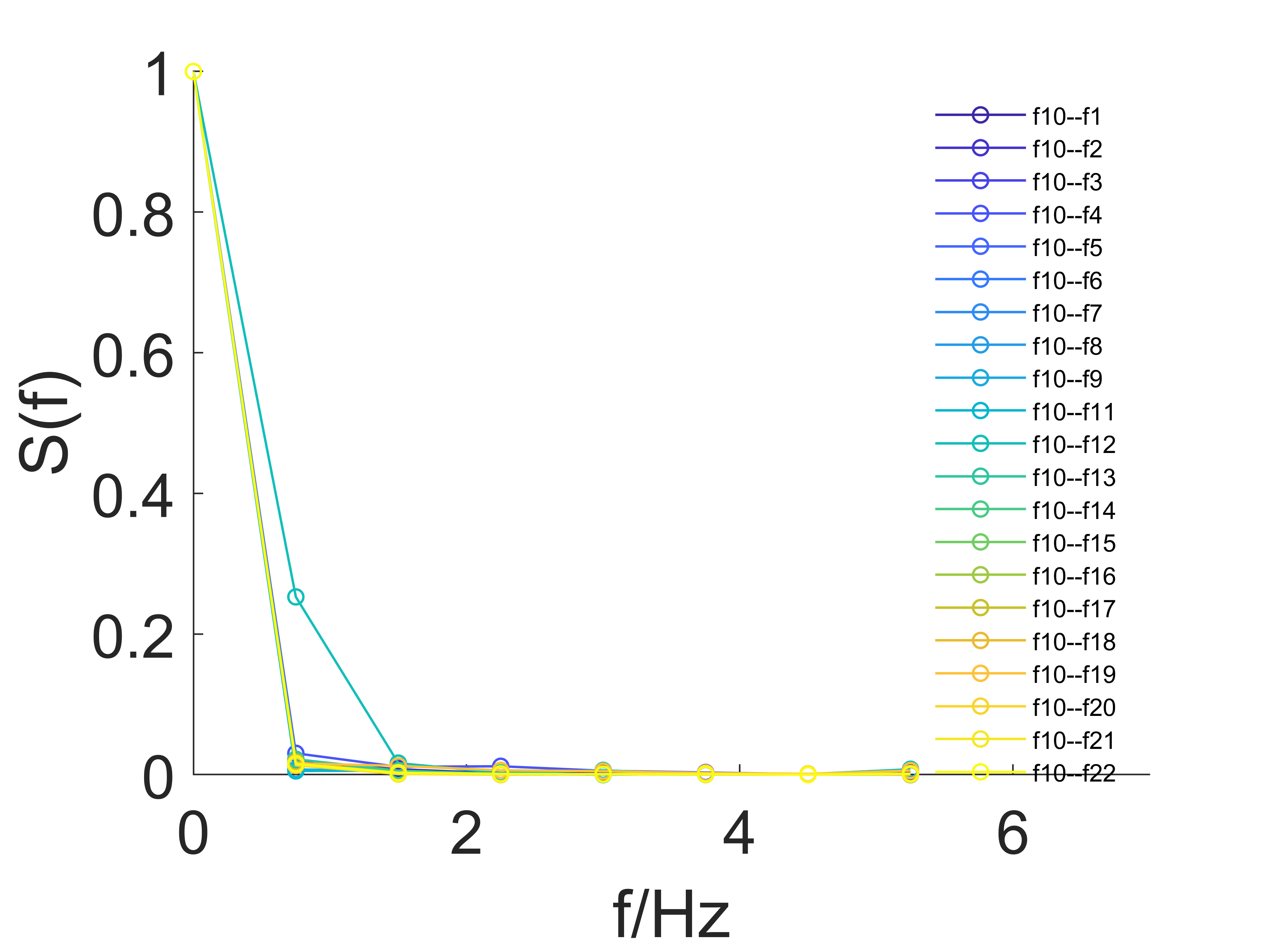}} 
\caption{$S(f)$ values between different signals.}
\label{fig.EcUND}
\vspace{-0.1cm}
\end{figure}

We select one time period [100.18s, 101.50s] which the proposed approach detects successfully and the LOF approach fails. For this time period, we calculate the $S(f)$ values between different signals for different frequency values. Fig. 3(a) shows the $S(f)$ values between $f$1 and other 20 signals except $f$10. Fig. 3(b) gives the $S(f)$ values between $f$10 and other 21 signals. Comparing them, we can see that $f$10 is very different with other signals in term of frequency magnitude for non-DC component. However, $f$1 and other signals (excluding $f$10) are much more similar. The dynamic change in time domain for the selected time period is not significant and the LOF approach is more sensitive to dynamic change and thus fails to detect it. The proposed approach also considers the frequency characteristic of data points and it is more accurate to quantify the similarity and differentiate different signals.

\subsection{Scenario 2: Performance under event condition}
The synchrophasor data recorded from a frequency event as shown by Fig. 4 is used. There are 20 frequency signals. 
%
%\begin{figure}[!htbp]
% \centering
% \vspace{-0.3cm}
%\subfloat[Added false data injection %]{\includegraphics[width=46mm]{Signals_add_data_spike_event2.png}}
%\subfloat[Identified false data injection  ]{\includegraphics[width=46mm]{Signals_identified_data_spike_event2.png}} 
%\caption{Frequency signals for Scenario 2.}
% \vspace{-0.3cm}
%\label{fig.EcUND}
%\end{figure}

%Considering false data injection caused by cyber attacks in 4 time periods %of 4 PMU signals as shown in Fig. 2(a). For each PMU signal among them, 5 %new data points replace the original data points within one time period. %The identified data points in 4 time periods of 4 signals using the %proposed approach are marked by red in Fig. 2(b). It verifies the %effectiveness of the proposed approach. 

Using the proposed approach, no low-quality data is detected. It indicates that the proposed approach will not cause wrong detection of low-quality data under event condition.

\subsection{Scenario 3: Detection of low-quality synchrophasor data that appears in the same time period of multiple regional PMU signals under event condition}
\begin{figure}[!htb]
\vspace{-0.2cm}
\begin{minipage}{0.23\textwidth}
 \centering
 \includegraphics[width=1.08\linewidth]{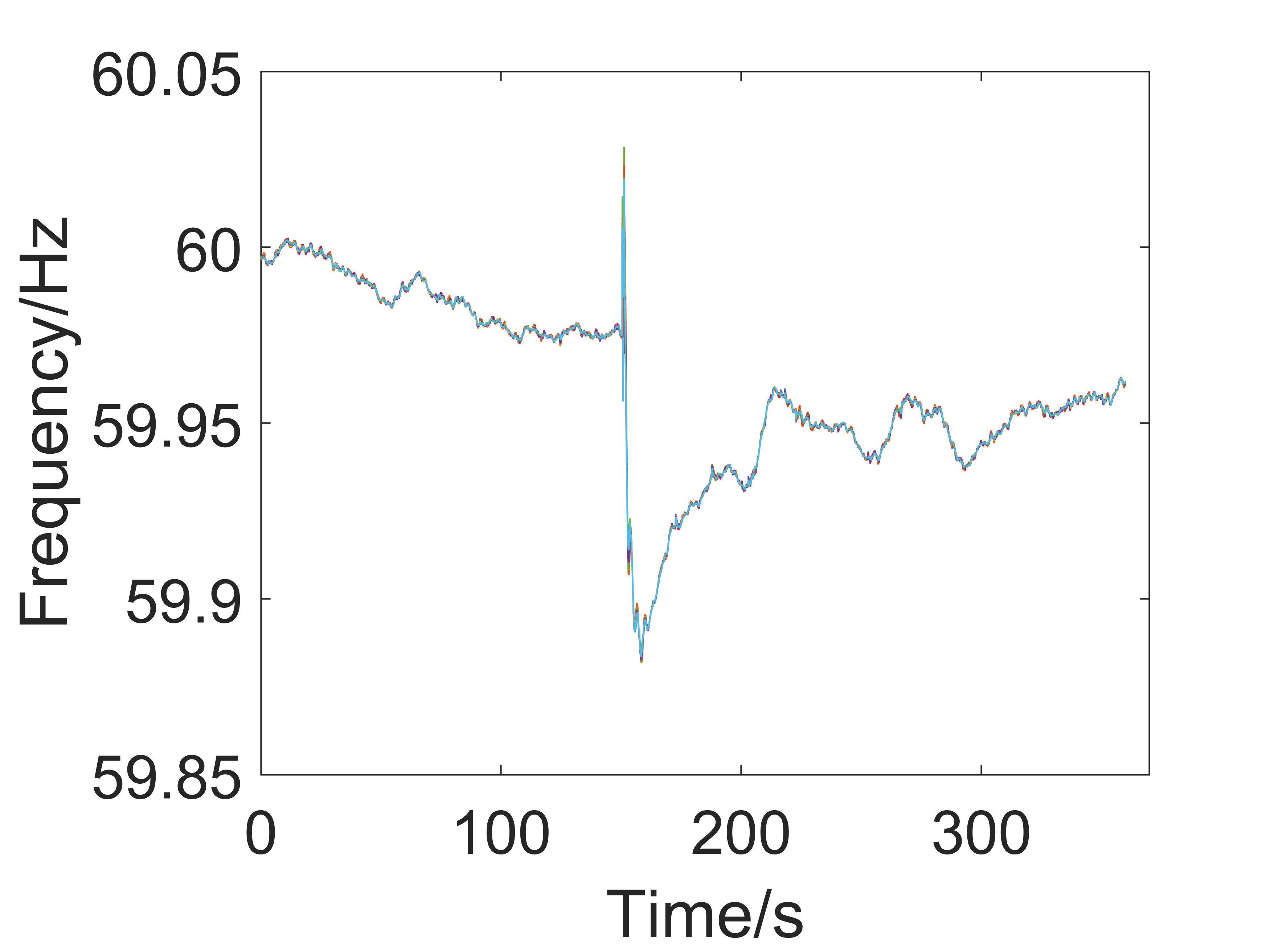}
 \caption{Signals for Scenario 2.}\label{Fig:Data1}
\end{minipage}\hfill
\begin {minipage}{0.23\textwidth}
 \centering
 \includegraphics[width=1.08\linewidth]{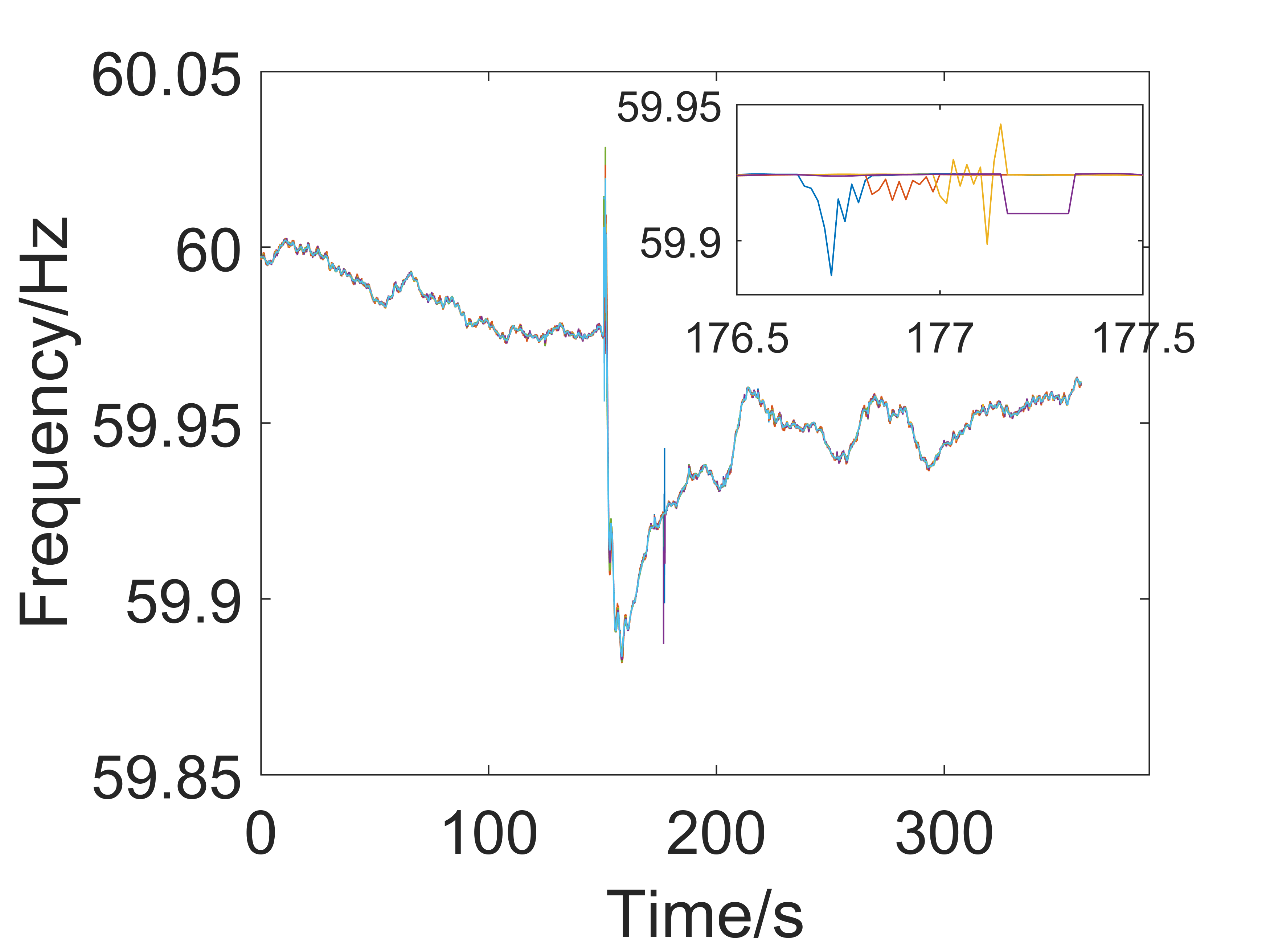}
 \caption{Signals for Scenario 3.}\label{Fig:Data2}
\end{minipage}
\vspace{-0.1cm}
\end{figure}

The LOF approach can detect false data injection that appears in the same time period of multiple regional PMU signals \cite{FA1}. This subsection verifies that the proposed approach is more effective to detect indistinguishable false data injection that appears in the same time period of multiple regional PMU signals when compared with the LOF approach. 

False data injection (10 data points) are introduced to the same time period of 4 signals($f$1,$f$2,$f$3,$f$4) shown in Fig. 5. 
\begin{figure}[!htbp]
\vspace{-0.15cm}
\centering
\subfloat[Proposed approach]{\includegraphics[width=46mm]{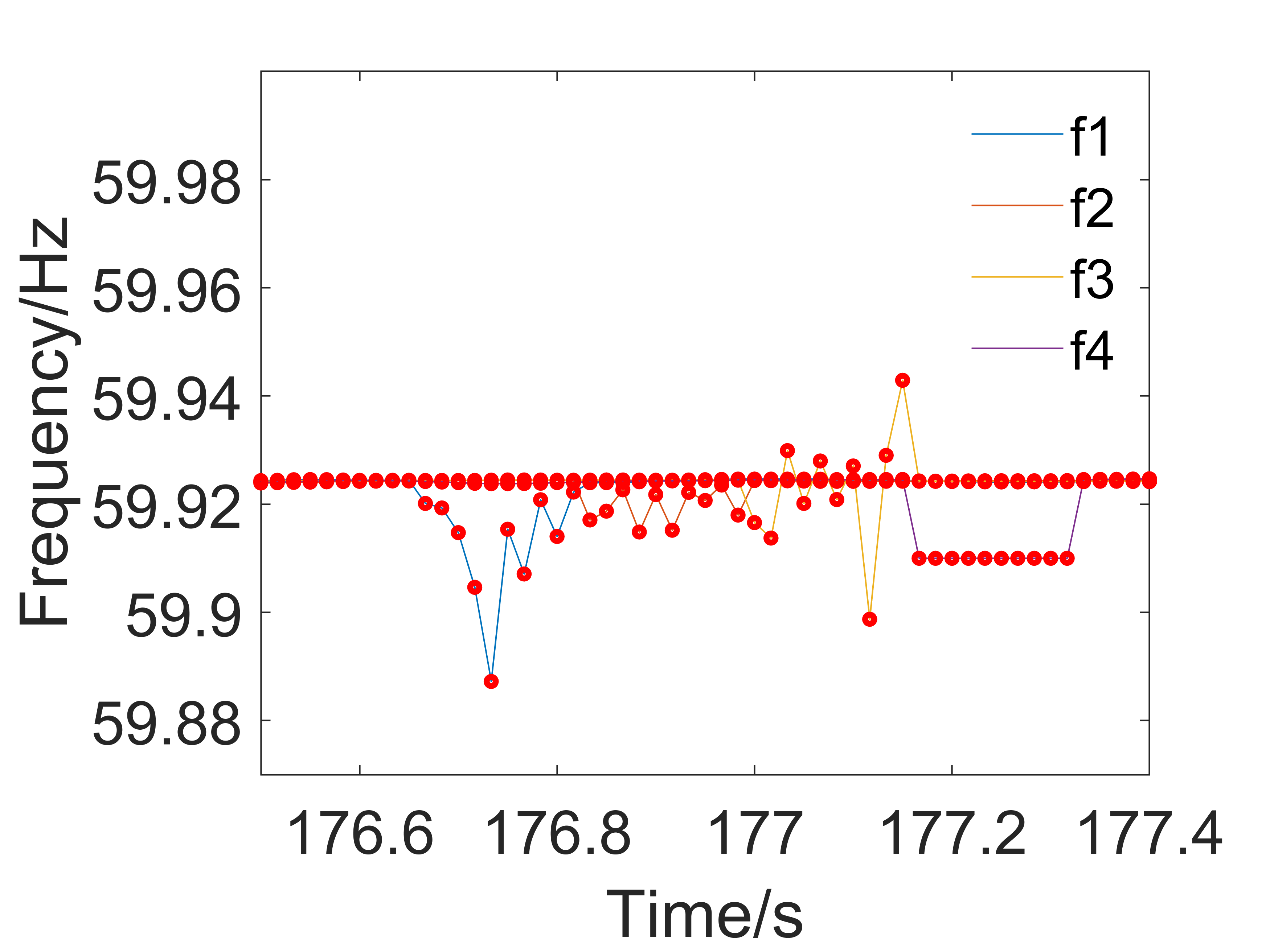}}
\subfloat[LOF approach ]{\includegraphics[width=46mm]{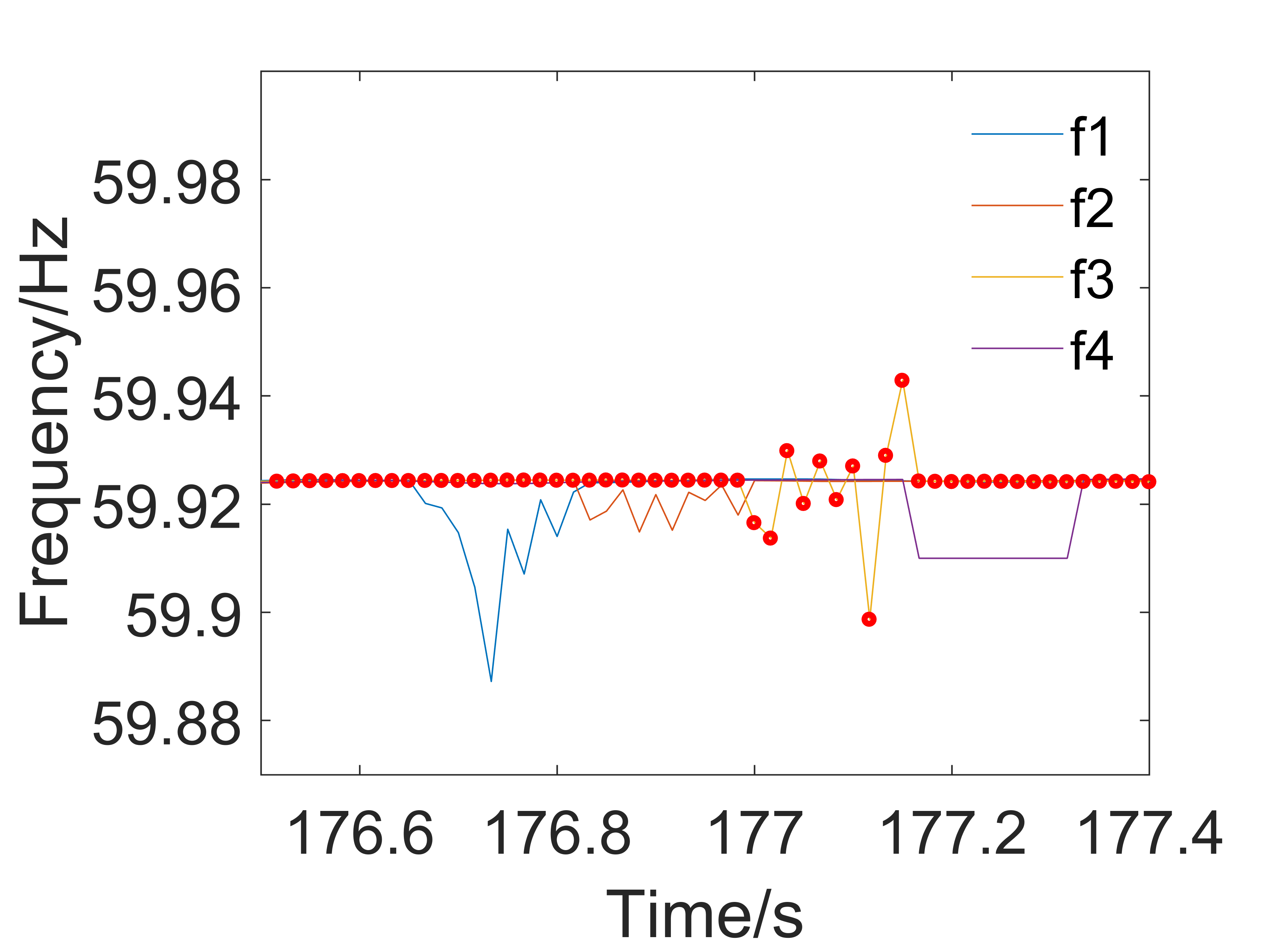}} 
\caption{Identified false data injection for Scenario 3.}
\label{fig.EcUND}
\end{figure}

The identified data points with low-quality data using the proposed and LOF approaches are marked by red in Fig. 6(a) and Fig. 6(b), respectively. Comparing them, we can find that the proposed approach detects the data points with false data injection for $f$1, $f$2, $f$3 and $f$4. However, the LOF approach only detect the data points with false data injection for $f$3.

1000 independent simulations are performed to randomly generate unobvious false data injection in terms of magnitude and compare the performance of two approaches. Table 1 gives the mean value of identified signals with false data injection in 1000 simulations. It verifies that the proposed approach is more capable to detect unobvious false data injection that appears in the same time period of multiple regional PMU signals since it also considers data characteristics in frequency domain to distinguish multiple signals. 
\begin{table}[!htbp]
\centering  
\caption{Mean Value of Identified Signals with False Data Injection}
\begin{tabular}{cc}
\hline
Approach & Value \\ \hline 
LOF Approach & 1.21 \\ 
Proposed Approach & 3.95  \\   \hline
\end{tabular}
\label{tab2}
\end{table}
\section{Conclusion}
This letter proposes a novel approach for detecting low-quality synchrophasor data under normal and event operating conditions. It utilizes the features of synchrophasor data in both time and frequency domains. The proposed approach is more effective to detect unobvious low-quality synchrophasor data such as random spikes and false data injection. It does not involve any offline study or training. Case studies for different scenarios verify the proposed approach.

\end{document}